\documentclass[journal]{IEEEtran}
\usepackage{cite,graphicx,amsmath,psfrag,bm}
\usepackage{psfrag}
\usepackage[usenames,dvipsnames]{xcolor}
\usepackage{mathabx}
\usepackage{amsbsy}
\usepackage{epstopdf}
\usepackage{subfigure}
\usepackage{setspace}
\usepackage{tikz}
\usetikzlibrary{matrix}
\usepackage{pstool}

\newcommand{\ve}[1]{\boldsymbol{#1}}

\begin{document}

\title{Simulation of Metasurfaces \\ in Finite Difference Techniques}

\author{Yousef Vahabzadeh, Karim Achouri
        and~Christophe~Caloz,~\IEEEmembership{Fellow,~IEEE}

\thanks{Y. Vahabzadeh, K. Achouri and C. Caloz are with the Department
of Electrical Engineering, $\acute{\mathrm{E}}$cole Polytechnique de Montr$\acute{\mathrm{e}}$al, Montr$\acute{\mathrm{e}}$al,
QC, H3T 1J4 Canada (e-mail: ).}
\thanks{}}

\markboth{}%
{Shell \MakeLowercase{\textit{et al.}}: Simulation of Zero-Thickness Electromagnetic Sheet Using Finite difference Technique}

\maketitle

\begin{abstract}
We introduce a rigorous and simple method for analyzing metasurfaces, modeled as zero-thickness electromagnetic sheets, in Finite Difference (FD) techniques. The method consists in describing the spatial discontinuity induced by the metasurface as a virtual structure, located between nodal rows of the Yee grid, using a finite difference version of Generalized Sheet Transition Conditions (GSTCs). In contrast to previously reported approaches, the proposed method can handle sheets exhibiting both electric and magnetic discontinuities, and represents therefore a fundamental contribution in computational electromagnetics. It is presented here in the framework of the FD Frequency Domain (FDFD) method but also applies to the FD Time Domain (FDTD) scheme. The theory is supported by five illustrative examples.
\end{abstract}


\IEEEpeerreviewmaketitle

\section{Introduction}

Metasurfaces are \emph{sub-wavelengthly thin} two-dimensional arrays of scattering particles~\cite{MM1,MM2}. They may be seen as dimensional reductions of three-dimensional metamaterials~\cite{liu, caloz, engeta} and functional extensions of frequency or polarization selective surfaces~\cite{munk,wu}. Compared to three-dimensional metamaterials, they feature lower loss, lighter weight and easier fabrication. Compared to frequency or polarization selective surfaces, they offer a much wider range applications, including phase and polarization transformers~\cite{ppc}, ultra-thin absorbers~\cite{absorber}, Faraday rotators~\cite{Kodera_APL_07_2011}, spatial waveguides~\cite{MetaWG}, generalized refractors~\cite{capasso1}, aberration-free lenses~\cite{ABBLens}, and ``spatial transistors''~\cite{Achouri_EPJAM_01_2016}, to cite a few.

The most general metasurfaces may be \emph{bianisotropic} and hence represent very complex spatial discontinuities~\cite{karim,Grbic,kuester}. In particular, they typically exhibit both electric and magnetic discontinuities~\cite{Keuster} in the Huygens regime, where particles with orthogonal and equal electric and magnetic dipole moments~\cite{SHuy} are used to suppress reflection~\cite{PFF_Hyu}. In contrast, many other two-dimensional structures, such as for instance graphene~\cite{Grise}, are monoisotropic or monoanisotropic\footnote{A bianisostropic is a medium whose constitutive relations are $\overrightarrow{D}=\overline{\overline{\epsilon}}\cdot\overrightarrow{E}+\overline{\overline{\xi}}\cdot\overrightarrow{H}$ and $\overrightarrow{B}=\overline{\overline{\zeta}}\cdot\overrightarrow{E}+\overline{\overline{\mu}}\cdot\overrightarrow{H}$. We call here ``monoanisotropic'' the particular case of such a medium when $\overline{\overline{\xi}}=\overline{\overline{\zeta}}=0$, and ``monoisotropic'' the further particular case where $\overline{\overline{\epsilon}}$ and $\overline{\overline{\mu}}$ reduce to scalars, $\epsilon$ and $\mu$.} and exhibit only magnetic discontinuity~\cite{GCaloz}. For this reason, metasurfaces may be regarded as the most general possible type of electromagnetic sheet.

Given their generality and richness, there is great motivation in developing efficient computational tools to analyze metasurfaces. FDTD~\cite{Susan} and FDFD~\cite{rumpf} are among the most popular numerical methods for solving Maxwell equations in the time and frequency domains, respectively, given their simplicity and ease of implementation, and they have already been applied to simulate some electromagnetic sheets. In~\cite{2DEG}, a two dimensional electron gas (2DEG) of zero thickness was studied by FDTD, while ~\cite{GRFDTD}, a more advanced FDTD scheme was used for analyzing graphene. However, none of these methods are applicable to a general metasurfaces, that may exhibit both magnetic and electric discontinuities, and in addition exhibit bianisotropy. To our knowledge, no simulation method has been reported yet to solve such a problem. The present paper fills up this gap by introducing a GSTC~\cite{idemen} treatment of the metasurface in an FDFD scheme, and may straightforwardly apply to FDTD as well. Our method is very simple and does not require modification of the FD equations elsewhere than in the vicinity of the metasurface.

The organization of the paper is as follows. Section~\ref{sec:rec_met_synth} recalls the GSTC metasurface synthesis equations. Section~\ref{sec:form_discont_prob} derives the 1D and 2D discontinuity equations and describes their implementation. In section~\ref{sec:ill_ex}, the proposed method is verified by several examples and the results are compared with COMSOL simulations of sub-wavelength thick metasurface. Finally, conclusions are provided in Sec.~\ref{sec:concl}.

\section{Recall of Metasurface Synthesis Equations}\label{sec:rec_met_synth}

\begin{figure}
\centering
\includegraphics[width=0.9\columnwidth]{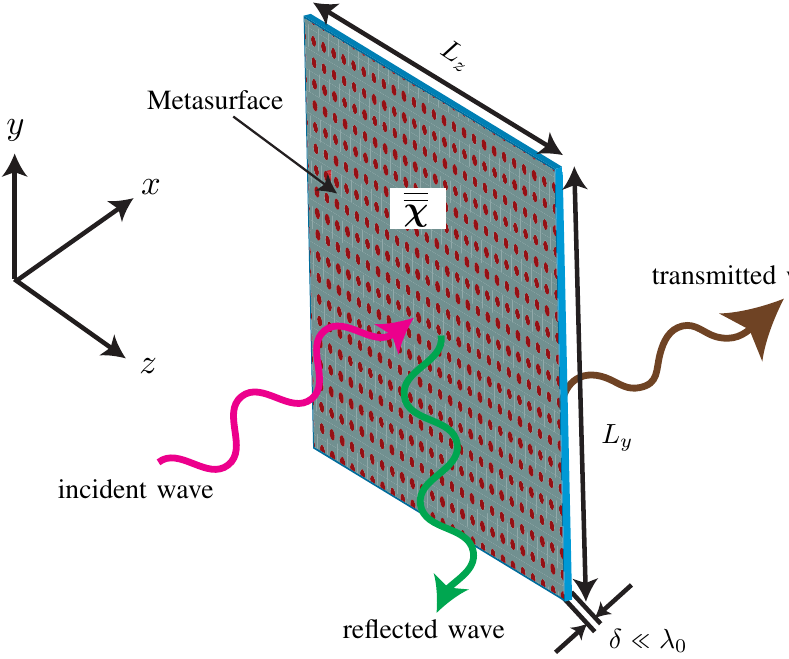}
  \caption{sheet discontinuity and cartesian coordinates system. The metasurface thickness, $\delta$, is much smaller than the operating wavelength, $\lambda_0$, and the metasurface will therefore be modeled as a zero-thickness sheet discontinuity.}
  \label{Fig:MSfig}
\end{figure}

In a metasurface perpendicular to the $x$ direction of a cartesian coordinate system and surrounded by two identical media, depicted in Fig.~\ref{Fig:MSfig}, the electric field, the magnetic fields and the polarization densities are related as~\cite{Keuster,idemen,karim}

\begin{subequations}
  \begin{align}\label{HPM}
   \hat{x}\times{\Delta{\overrightarrow{H}}}&=j\omega \overrightarrow{P}_{\Vert}-\hat{x} \times \nabla_{\Vert}M_x, \\\label{EPM}
   {\Delta{\overrightarrow{E}}}\times\hat{x}&=j\omega\mu\overrightarrow{M}_{\Vert}-\nabla_{\Vert}\left( \frac{P_x}{\varepsilon} \right)\times\hat{x},\\
   \hat{x}\cdot\Delta \overrightarrow{D}&=-\nabla \cdot\overrightarrow{P}_{\Vert},\\
   \hat{x}\cdot\Delta \overrightarrow{B}&=-\mu\nabla\cdot\overrightarrow{M}_{\Vert},
  \end{align}
\end{subequations}
where $\nabla_{\Vert}=\frac{\partial}{\partial{z}}\hat{z}+\frac{\partial}{\partial{y}}\hat{y}$, $\overrightarrow{M}$ is the magnetic polarization density, $\overrightarrow{P}$ is the electric polarization density and $\Delta\overrightarrow{\psi}=\overrightarrow{\psi}^{\textrm{tr}}-(\overrightarrow{\psi}^{\textrm{ref}}+\overrightarrow{\psi}^{\textrm{inc}})$ with $\textrm{tr}, \textrm{ref}$ and $\textrm{inc}$ denoting the transmitted, reflected and incident waves, respectively.

The polarization densities may be expressed as
\begin{subequations}
  \begin{align}\label{PEH}
   \overrightarrow{P}&=\epsilon_0\overline{\overline{\chi}}_{\textrm{ee}}\overrightarrow{E}_{\textrm{av}}+\overline{\overline{\chi}}_{\textrm{em}}\sqrt{\epsilon_0\mu_0}\overrightarrow{H}_{\textrm{av}},\\\label{MHE}
   \overrightarrow{M}&=\overline{\overline{\chi}}_{\textrm{mm}}\overrightarrow{H}_{\textrm{av}}+\overline{\overline{\chi}}_{\textrm{me}}\sqrt{\frac{\epsilon_0}{\mu_0}}\overrightarrow{E}_{\textrm{av}},
  \end{align}
\end{subequations}
where $\overline{\overline{\chi}}_{\textrm{ee}}, \overline{\overline{\chi}}_{\textrm{mm}}, \overline{\overline{\chi}}_{\textrm{em}}$ and $\overline{\overline{\chi}}_{\textrm{me}}$ are the electric/magnetic (first e/m subscripts) susceptibility tensors describing the response to electric/magnetic (second e/m subscripts) excitations, and where the subscript av denotes the average of the fields on both sides of the metasurface, $\overrightarrow{\psi}_\text{av}=
[(\overrightarrow{\psi}^{\textrm{inc}}+\overrightarrow{\psi}^{\textrm{ref}})+\overrightarrow{\psi}^{\textrm{inc}}]/2$.
Without any loss of generality as far as the proposed method is concerned, we assume here that $P_x=M_x=0$, in which case substituting~\eqref{PEH} and~\eqref{MHE} into~\eqref{HPM} and~\eqref{EPM}, respectively, yields the following simple linear system of equations~\cite{karim}:
\begin{subequations}\label{GSTC}
  \begin{align}\label{GSTC1}
  \left(
  \begin{array}{c}
    -\Delta H_z \\
    \Delta H_y \\
  \end{array}
\right) =j\omega\varepsilon_0 &\left(
                        \begin{array}{cc}
                          \chi _{\textrm{ee}}^{yy} & \chi _{\textrm{ee}}^{yz} \\
                          \chi _{\textrm{ee}}^{zy} & \chi _{\textrm{ee}}^{zz} \\
                        \end{array}
                      \right)
                      \left(
                        \begin{array}{c}
                          E_{y,\textrm{av}} \\
                          E_{z,\textrm{av}} \\
                        \end{array}
                      \right)\\\notag
                      &+j\omega\sqrt{\varepsilon_0 \mu_0} \left(
                                                           \begin{array}{cc}
                                                             \chi _{\textrm{em}}^{yy} & \chi _{\textrm{em}}^{yz} \\
                                                             \chi _{\textrm{em}}^{zy} & \chi _{\textrm{em}}^{zz} \\
                                                           \end{array}
                                                         \right)
                                                         \left(
                                                           \begin{array}{c}
                                                             H_{y,\textrm{av}} \\
                                                             H_{z,\textrm{av}} \\
                                                           \end{array}
                                                         \right),
  \end{align}
  \begin{align}\label{GSTC2}
  \left(
  \begin{array}{c}
    -\Delta E_y \\
    \Delta E_z \\
  \end{array}
\right) =j\omega\mu_0 &\left(
                        \begin{array}{cc}
                          \chi _{\textrm{mm}}^{zz} & \chi _{\textrm{mm}}^{zy} \\
                          \chi _{\textrm{mm}}^{yz} & \chi _{\textrm{mm}}^{yy} \\
                        \end{array}
                      \right)
                      \left(
                        \begin{array}{c}
                          H_{z,\textrm{av}} \\
                          H_{y,\textrm{av}} \\
                        \end{array}
                      \right)\\\notag
                      &+j\omega\sqrt{\varepsilon_0 \mu_0} \left(
                                                           \begin{array}{cc}
                                                             \chi_{\textrm{me}}^{zz} & \chi _{\textrm{me}}^{zy} \\
                                                             \chi_{\textrm{me}}^{yz} & \chi _{\textrm{me}}^{yy} \\
                                                           \end{array}
                                                         \right)
                                                         \left(
                                                           \begin{array}{c}
                                                             E_{z,\textrm{av}} \\
                                                             E_{y,\textrm{av}} \\
                                                           \end{array}
                                                         \right).
\end{align}
\end{subequations}

Representing generally a system of 4 equations in 16 unknowns, this matrix system reveals that a fully bianisotropic metasurface, with $P_x=M_x=0$, allows up to 4 wave transformation, i.e. 4 independent ($\psi^{\textrm{inc}},\psi^{\textrm{ref}},\psi^{\textrm{tr}})$ triplets~\cite{karim}. In the following formulations and discussions, we will consider only one triplet to avoid lengthy equations and heavy notation. However, the following procedure allows straightforward extension to the case of multiple transformations~\cite{karim}.

\section{GSTC-Based Method}\label{sec:form_discont_prob}

\subsection{1D Case}

\begin{figure}
\centering
\includegraphics[width=1\columnwidth]{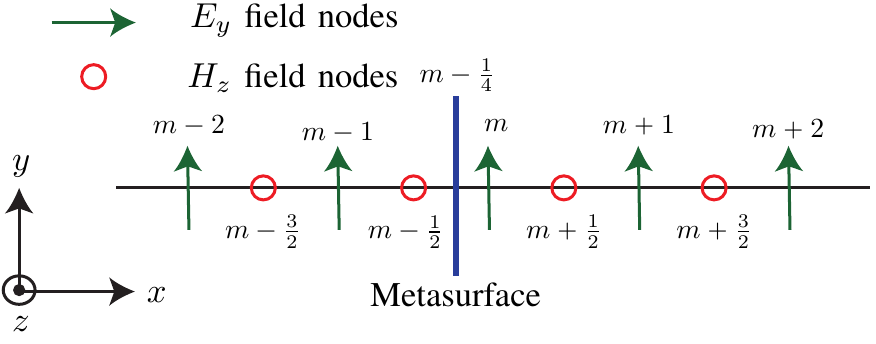}
  \caption{Position of the sheet in the 1D Yee grid. The $E_y$ and $H_z$ nodes are located at integer and half integer points, respectively. The incident wave impinges on the sheet from the left side and the transmitted wave exits the sheet at the right side.}
  \label{Fig:1DFDFD}
\end{figure}

We first consider the 1D problem, with propagation occurring in the $x$ direction, and $H_z$ and $E_y$ being the only non-zero field components, as shown in Fig.~\ref{Fig:1DFDFD}. In this case, the coupling susceptibilities are zero $\left(\overline{\overline{\chi}}_{\textrm{em}}=\overline{\overline{\chi}}_{\textrm{me}}=0\right)$, and~\eqref{GSTC} reduces to

\begin{subequations}\label{1DGSTC}
  \begin{align}\label{1DGSTC1}
   -\Delta H_z&=j\omega\varepsilon_0\chi_{\textrm{ee}}^{yy}E_{y,\textrm{av}},\\\label{1DGSTC2}
   -\Delta E_y&=j\omega\mu_0\chi_{\textrm{mm}}^{zz}H_{z,\textrm{av}}.
  \end{align}
\end{subequations}

From~\cite{rumpf}, the corresponding standard FDFD equations read
 \begin{subequations}\label{1DFDeq}
   \begin{align}\label{1DFDeq1}
     \frac{E_y^{i+1}-E_y^{i}}{\Delta x} =-j\omega\mu_0\mu_{zz}^{i+\frac{1}{2}}H_z^{i+\frac{1}{2}}  \quad &\textrm{or} \quad \ve{D}_\textrm{e}^x\ve{E}_y=\ve{\mu}_{zz}\ve{H}_z,\\\label{1DFDeq2}
     -\frac{H_z^{i+\frac{1}{2}}-H_z^{i-\frac{1}{2}}}{\Delta x}=j\omega\varepsilon_0\varepsilon_{yy}^{i}E_y^{i} \quad &\textrm{or} \quad -\ve{D}_\textrm{h}^x\ve{H}_z=\ve{\varepsilon}_{yy}\ve{E}_y,
   \end{align}
 \end{subequations}
where $\ve{D}_\textrm{e}^x$ and $\ve{D}_\textrm{h}^x$ are the differential operator matrices, $\ve{\mu}_{zz}$ and $\ve{\varepsilon}_{yy}$ are the material matrices, and $\ve{H}_z$ and $\ve{E}_y$ are the field vectors given in App.~\ref{sec:APP1}.

According to~\eqref{1DGSTC}, the metasurface represents a locus where both the electric field \emph{and} the magnetic field are discontinuous, since $\Delta E_y\neq 0$ and $\Delta H_z\neq 0$, where we recall that $\Delta$ here represents a discontinuity across the zero-thickness sheet modeling the metasurface. Consequently, the metasurface can be placed neither at an electric field grid node nor at a magnetic field grid node since such nodes would fail to account for the corresponding discontinuity. Therefore, we treat the metasurface as a virtual structure and place it \emph{between} the $(m-1)^{\textrm{th}}$ and $m^{\textrm{th}}$ nodes of the Yee grid, as shown in Fig.~\ref{Fig:1DFDFD}.

The FDFD equations~\eqref{1DFDeq} can be employed everywhere except at grid nodes surrounding the metasurface discontinuity, where a special GSTC treatment will be applied. For these nodes, consider the update equation~\eqref{1DFDeq1}. Updating $H_z^{m+\frac{1}{2}}$ involves $E_y^m$ and $E_y^{m+1}$. This is not a problem since both of these nodes are located at the same (right) side of the discontinuity. Therefore, the standard FDFD equation~\eqref{1DFDeq1} can be employed for updating $H_z^{m+\frac{1}{2}}$. In contrast, updating $H_z^{m-\frac{1}{2}}$ involves $E_y^{m-1}$ and $E_y^m$ with the former being located on the other (left) side of the metasurface, thus requiring specific account of the related discontinuity. For this case, we propose to employ the GSTC relation~\eqref{1DGSTC2} instead of the standard FDFD~\eqref{1DFDeq1}, which leads to
\begin{equation}\label{C1}
   -E_y^{m}+E_y^{m-1}
   =\frac{j\omega\mu_0\chi_{\textrm{mm}}^{zz,(m-1/4)}}{2}\left(H_z^{m+\frac{1}{2}}+H_z^{m-\frac{1}{2}}\right).
\end{equation}
This relation properly accounts for the discontinuity, involving the appropriate metasurface susceptibility.

Let us now consider the other update equation, Eq.~\eqref{1DFDeq2}, and apply the same logics to it. Updating $E_y^{m-1}$ involves $H_z^{m-\frac{1}{2}}$ and $H_z^{m-\frac{3}{2}}$, which are both located at the same (left) side of the discontinuity. So, Eq.~\eqref{1DFDeq2} is still applicable. The situation is different for $E_y^m$, since $H_z^{m+\frac{1}{2}}$ is located at the other (right) side of the discontinuity. For this case, Eq.~\eqref{1DFDeq2} is replaced by~\eqref{1DGSTC1}, which gives
\begin{equation}\label{c2}
  -H_z^{m+\frac{1}{2}}+H_z^{m-\frac{1}{2}}
  =\frac{j\omega\varepsilon_0\chi_{\textrm{ee}}^{yy,(m-1/4)}}{2}\left(E_y^{m-1}+E_y^{m}\right),
\end{equation}
properly accounting for the discontinuity.

The implementation of~\eqref{C1} requires adjustment of the $\ve{D}_\textrm{e}^x$ and $\ve{\mu}_{zz}$ matrices, given in App.~\ref{sec:APP1}. However, the modification is minor and easy to implement. In~\eqref{Dez}, only the $(m-1)^{\textrm{th}}$ row needs to be altered: in this row all the entries must be set to zero except the $m^{\textrm{th}}$ and $\left(m-1\right)^{\textrm{th}}$ entries that become $\ve{D}_\textrm{e}^x(m-1,m-1)=1$ and $\ve{D}_\textrm{e}^x(m-1,m)=-1$; moreover, in~\eqref{muhz}, one sets $\ve{\mu}_{zz}\left(m-1,m\right)=\ve{\mu}_{zz}\left(m-1,m-1\right)=\frac{j\omega\mu_0\chi_\textrm{mm}^{zz}}{2}$, and all the other entries to zero. Similarly, the implementation of~\eqref{c2} requires modifying the $\ve{D}_\textrm{h}^x$ and $\ve{\varepsilon}_{yy}$ matrices. In~\eqref{Dhz}, only the $m^{\textrm{th}}$ row is altered: in this row, all the entries are set to zero except for $\ve{D}_\textrm{h}^x(m,m)=1$ and $\ve{D}_\textrm{h}^x(m,m-1)=-1$ while in the $\ve{\varepsilon}_{yy}$ matrix, $\ve{\varepsilon}_{yy}\left(m,m\right)=\ve{\varepsilon}_{yy}\left(m,m-1\right)=\frac{j\omega\varepsilon_0\chi_\textrm{ee}^{yy}}{2}$ with the other entries of the $m^{\textrm{th}}$ row being set to zero.

\subsection{2D Case}

\begin{figure}
\centering
\includegraphics[width=1\columnwidth]{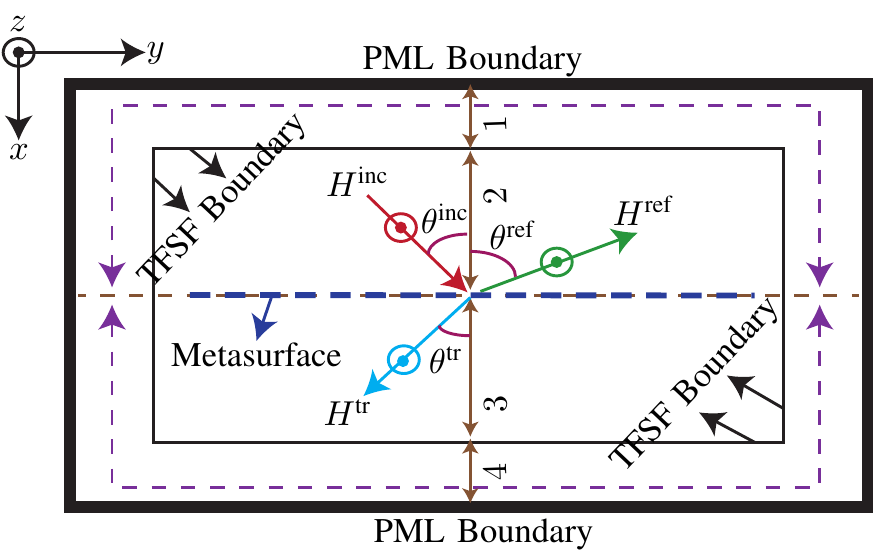}
\caption{Different regions in the FDFD computational domain and position of the metasurface in the total field region. The numbers $1, 2, 3$ and $4$ refer to the regions of the reflected field, the sum of the incident and reflected fields, the transmitted field and the transmitted field minus incident field. The Total-Field Scattered-Field, TFSF, boundary separates total-field and scattered-field regions. \label{fig:TFSFfig}}
\end{figure}

The extension of the method from 1D to 2D is straightforward. However, since a general treatment of Eqs.~\eqref{GSTC} leads to lengthy equations, we will restrict here our attention to the simple -- but most useful -- case of TE$_z$ polarization, illustrated in Fig.~\ref{fig:TFSFfig} and where the only nonzero field components are $E_x$, $E_y$ and $H_z$. In this case, Eqs.~\eqref{GSTC} reduce to

 \begin{subequations}\label{TE}
   \begin{align}\label{TE1}
     -\Delta H_z&=j\omega\varepsilon_0\chi_{\textrm{ee}}^{yy}E_{y,\textrm{av}}+jk_0\chi_{\textrm{em}}^{yz}H_{z,\textrm{av}},\\\label{TE2}
     -\Delta E_y&=j\omega\mu_0\chi_{\textrm{mm}}^{zz}H_{z,\textrm{av}}+jk_0\chi_{\textrm{me}}^{zy}E_{y,\textrm{av}}.
   \end{align}
 \end{subequations}

From~\cite{rumpf}, the corresponding standard FDFD equations read
  \begin{subequations}\label{2DFDeq}
   \begin{align}\label{2DFDeq1}
     &\frac{E_y^{i+1,j+\frac{1}{2}}-E_y^{i,j+\frac{1}{2}}}{\Delta x}-\frac{E_x^{i+\frac{1}{2},j+1}-E_x^{i+\frac{1}{2},j}}{\Delta y} \\\notag &=-j\omega\mu_0\mu_{zz}^{i+\frac{1}{2},j+\frac{1}{2}}H_z^{i+\frac{1}{2},j+\frac{1}{2}}\quad \textrm{or}\quad \ve{D}_e^x\ve{E}_y-\ve{D}_e^y\ve{E}_x=\ve{\mu}_{zz}\ve{H}_z, \\\label{2DFDeq2}
     &-\frac{H_z^{i+\frac{1}{2},j+\frac{1}{2}}-H_z^{i-\frac{1}{2},j+\frac{1}{2}}}{\Delta x} =j\omega\varepsilon_0\varepsilon_{yy}^{i,j+\frac{1}{2}}E_y^{i,j+\frac{1}{2}}\\\notag &\qquad \textrm{or}\quad -\ve{D}_h^x\ve{H}_z=\ve{\varepsilon}_{yy}\ve{E}_y, \\\label{2DFDeq3}
     &\frac{H_z^{i+\frac{1}{2},j+\frac{1}{2}}-H_z^{i+\frac{1}{2},j-\frac{1}{2}}}{\Delta y}
     =j\omega\varepsilon_0\varepsilon_{xx}^{i+\frac{1}{2},j}E_x^{i+\frac{1}{2},j}\\\notag &\qquad \textrm{or}\quad \ve{D}_h^y\ve{H}_z=\ve{\varepsilon}_{xx}\ve{E}_x.
   \end{align}
 \end{subequations}

\begin{figure}
\includegraphics[width=0.9\columnwidth]{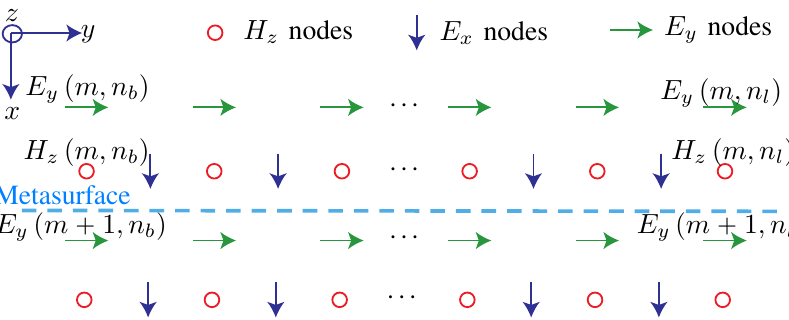}
\caption{Position of the metasurface in the 2D Yee grid, between the $H_z$ and $E_y$ field nodes at $i=m$ from $j=n_b$ through $j=n_l$ in the $y$ direction. As in Figs.~\ref{Fig:MSfig} and Fig.~\ref{fig:TFSFfig}, the $x$ axis is normal to the metasurface. $H_z$ is measured at half integer $x$ and $y$ points, $E_y$ is measured at half integer $y$ points but full integer $x$ points and $E_x$ is measured at half integer $x$ but full integer $y$ points. The numbers in parenthesis refer to the cell numbers. For instance, $E_y(m,n)$ represents the $m^{\textrm{th}}$ and $n^{\textrm{th}}$ cell in the $x$ and $y$ directions, respectively.}\label{fig:MYEE}
\end{figure}

As in 1D, the FDFD equations~\eqref{2DFDeq} are applied everywhere in space except at some nodes around the metasurface discontinuity, where~\eqref{TE} will be used instead. Figure~\ref{fig:TFSFfig} shows the computational box and its usual regions~\cite{rumpf}. The metasurface is completely immersed within the total field region for generating maximal information on scattering. Similar to the 1D case, the metasurface is considered as a virtual structure located between the $m^{\textrm{th}}$ and $(m+1)^{\textrm{th}}$ Yee grid nodes and, more specifically, between $E_y(m+1,n_b:n_l)$ and $H_z(m,n_b:n_l)$, as shown in Fig.~\ref{fig:MYEE}.

For the nodes around the metasurface, consider~\eqref{2DFDeq2}, where $H_z(m,n_b:n_l)$ and $H_z(m-1,n_b:n_l)$ are required to update $E_y(m,n_b:n_l)$. These two $H_z$ terms are located at the same (top) side of the metasurface, and therefore the standard FDFD equation~\eqref{2DFDeq2} are still applicable. In contrast, updating $E_y(m+1,n_b:n_l)$ involves $H_z(m+1,n_b:n_l)$ that is located on other (bottom) side of the discontinuity. To properly account for that discontinuity,~\eqref{2DFDeq2} is replaced by~\eqref{TE1}, which explicitly reads
\begin{align}
  -H_z^{\textrm{tr}}+H_z^{\textrm{inc}}+H_z^{\textrm{ref}}=&j\omega\varepsilon_0\chi_{\textrm{ee}}^{yy}\frac{E_y^{\textrm{inc}}+E_y^{\textrm{tr}}+E_y^{\textrm{ref}}}{2}\\\notag
  &+jk_0\chi_{\textrm{em}}^{yz}\frac{H_z^{\textrm{inc}}+H_z^{\textrm{tr}}+H_z^{\textrm{ref}}}{2}.
\end{align}
Discretizing this relation provides for the 2D counterpart of~\eqref{c2}
\begin{subequations}\label{2DTEEy}
\begin{equation}
\begin{split}
  &H_z^{m+\frac{1}{2},j+\frac{1}{2}}\left(1-\alpha_1(j)\right)-H_z^{m+\frac{3}{2},j+\frac{1}{2}}\left(1+\alpha_1(j)\right)\\
  &\quad=\frac{ j\omega\varepsilon_0\chi_{\textrm{ee}}^{yy,\left(m+\frac{1}{4},j+\frac{1}{2}\right)}}{2}\left( E_y^{m,j+\frac{1}{2}}+E_y^{m+1,j+\frac{1}{2}} \right),
\end{split}
\end{equation}
with \begin{equation}
\alpha_1(j)=\frac{jk_0\chi_{\textrm{em}}^{yz,\left(m+\frac{3}{4},j+\frac{1}{2}\right)}}{2}.
\end{equation}
\end{subequations}

The same procedure is used to update $H_z$ at the nodes around the metasurface. For $H_z(m+1,n_b:n_l)$, the standard FDFD equation~\eqref{2DFDeq1} is used, while for $H_z(m,n_b:n_l)$, Eq.~\eqref{2DFDeq1} is replaced by~\eqref{TE2}, which yields
\begin{equation}
\begin{split}
  -E_y^{\textrm{tr}}+E_y^{\textrm{inc}}+&E_y^{\textrm{ref}}=j\omega\mu_0\chi_{\textrm{mm}}^{zz}\frac{\left(H_z^{\textrm{inc}}+H_z^{\textrm{ref}}+H_z^{\textrm{tr}}\right)}{2}\\
  &+j\omega\sqrt{\varepsilon_0\mu_0}\chi_{\textrm{me}}^{zy}\frac{\left(E_y^{\textrm{inc}}+E_y^{\textrm{ref}}+E_y^{\textrm{tr}}\right)}{2}.
\end{split}
\end{equation}
After discretization, this relation becomes the 2D counterpart of~\eqref{C1}, reading
\begin{subequations}\label{2DFDHz}
\begin{equation}
\begin{split}
  &E_y^{m,j+\frac{1}{2}}\left(1-\alpha_2(j)\right)-E_y^{m+1,j+\frac{1}{2}}\left(1+\alpha_2(j)\right)\\
  &\quad=\frac{j\omega\mu_0\chi_{\textrm{mm}}^{zy,\left( m+\frac{3}{4},j+\frac{1}{2}\right)}}{2}\left(H_z^{m+\frac{1}{2},j+\frac{1}{2}}+H_z^{m+\frac{3}{2},j+\frac{1}{2}}\right),
\end{split}
\end{equation}
with
\begin{equation}
\alpha_2(j)=\frac{jk_0\chi_{\textrm{me}}^{zy,\left(m+\frac{3}{4},j+\frac{1}{2}\right)}}{2}.
\end{equation}
\end{subequations}

The implementation of~\eqref{2DTEEy} requires the following changes in the operator $\ve{D}_h^x$ and in the matrix $\ve{\varepsilon}_{yy}$, given in App.~\ref{sec:APP1}: $\ve{D}_h^x(i,i)=1+\alpha_1(j)$ and $\ve{D}_h^x(i,i-1)=-1+\alpha_1(j)$ with $i=m+1+(j-1)n_x$ with $j=n_b:n_l$ and where $n_x$ is number of cells in the $x$ direction in Fig.~\ref{fig:MYEE}; $\ve{\varepsilon}_{yy}(i,i)=\ve{\varepsilon}_{yy}(i,i-1)=j\omega\varepsilon_0\frac{\chi_{\textrm{ee}}^{yy}(j)}{2}$. Similarly, the implementation of~\eqref{2DFDHz} requires the following changes in the operator $\ve{D}_\textrm{e}^x$ and $\ve{D}_\textrm{e}^y$ and in the matrix $\ve{\mu}_{zz}$: $\ve{D}_\textrm{e}^x(i,i)=1-\alpha_2$, $\ve{D}_\textrm{e}^x(i,i+1)=-1-\alpha_2$ with $i=m+(j-1)n_x$ where $j=n_b:n_l$, and $\ve{\mu}_{zz}(i,i)=\ve{\mu}_{zz}(i,i+1)=\frac{j\omega\mu_0\chi_{\textrm{mm}}^{zz}(j)}{2}$ and $\ve{D}_\textrm{e}^y(i,:)=0$.

As a general rule, simulating a metasurface discontinuity with the proposed method requires replacing the standard FDFD equations with a FD difference version of GSTCs at the location of the metasurface. This replacement results in modified differential operators and material matrices.

\section{Illustrative Examples}\label{sec:ill_ex}

In the forthcoming five examples, the dimensions of the problem in the $x$ and $y$ directions are $20\lambda_0$ and $30\lambda_0$, respectively, with resolution (number of cells per wavelength) equal to~30. The simulation frequency is $f=10$~GHz and the width of the PML layers is 30~cells at all sides. The surrounding medium is free space. The reflection and transmission coefficients are noted $R$ and $T$, respectively.

\subsection{1D Examples}

\begin{figure}
\centering
\includegraphics[width=1\columnwidth]{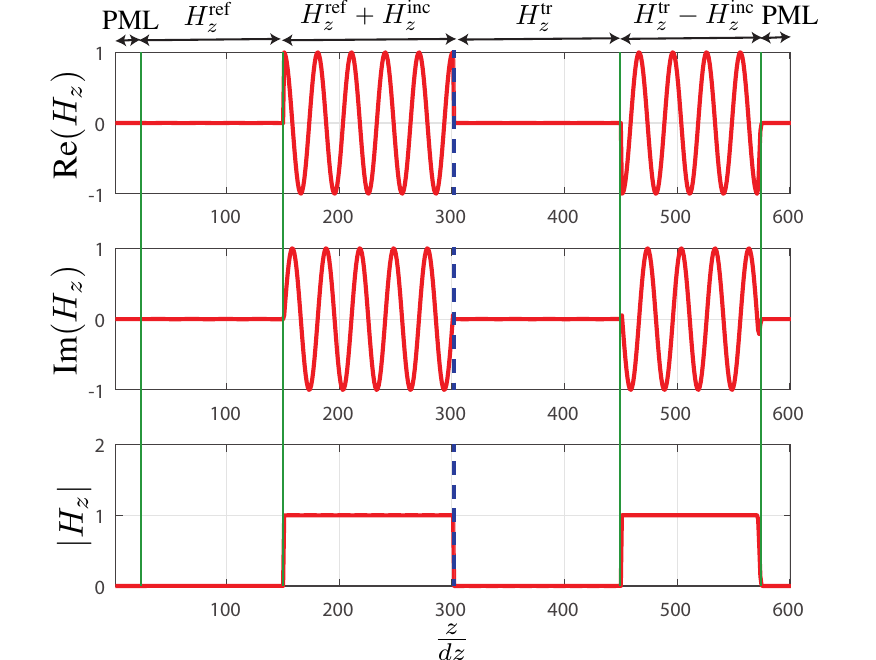}
\caption{Example~1 (1D): FDFD-GSTC results for a fully absorbing metasurface ($R=T=0$) with normally incident wave. The different computational regions are indicated at the top of the figure and the metasurface is located at $\frac{z}{dz}=300$, as indicated by the dashed line.}\label{Fig:1DAbsorber}
\end{figure}

The first 1D example is a fully absorbing metasurface, for which $R=T=0$. According to~\eqref{1DGSTC}, the corresponding synthesis metasurface susceptibilities are
\begin{subequations}
\begin{align}
\chi_{\textrm{ee}}^{yy}&=\frac{2}{j\omega\varepsilon_0}\frac{H_z^{\textrm{inc}}}{E_y^{\textrm{inc}}},\\ \chi_{\textrm{mm}}^{zz}&=\frac{2}{j\omega\mu_0}\frac{E_y^{\textrm{inc}}}{H_z^{\textrm{inc}}}.
\end{align}
\end{subequations}
The FDFD-GSTC results are presented in Fig.~\ref{Fig:1DAbsorber}. As expected, the metasurface is essentially absorptive, with a relative error in the order of $10^{-3}$.

\begin{figure}
\centering
\includegraphics[width=1\columnwidth]{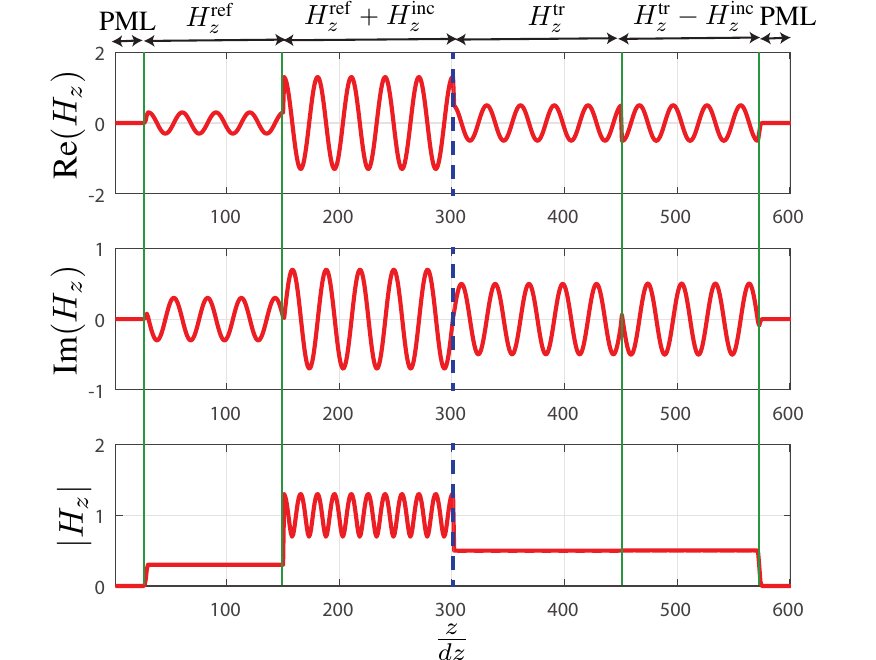}
  \caption{Example 2 (1D): FDFD-GSTC results for a metasurface with $R=0.3$ and $T=0.5$ (absorption $1-(R+T)=0.2$) with normally incident wave. Same conventions as in Fig.~\ref{Fig:1DAbsorber}.}\label{FIG:1DRT}
\end{figure}

The second 1D example is a metasurface involving reflection, transmission and absorption. In this case, the synthesis metasurface susceptibilities in~\eqref{1DGSTC} are found as
\begin{subequations}
\begin{align}
\chi_{\textrm{ee}}^{yy}=-\frac{2}{j\omega\varepsilon_0}\frac{\left(-T+1+R\right)H_z^{\textrm{inc}}}{\left(1+T-R\right)E_y^{\textrm{inc}}},\\ \chi_{\textrm{mm}}^{zz}=-\frac{2}{j\omega\mu_0}\frac{\left(-T+1-R\right)E_y^{\textrm{inc}}}{\left(1+R+T\right)H_z^{\textrm{inc}}}.
\end{align}
\end{subequations}
The results are shown in Fig.~\ref{FIG:1DRT} for $R=0.3$ and $T=0.5$. For the computed reflected wave, $0.300184 \leq|H_z|\leq 0.300675$, and for the transmitted wave, $0.497987 \leq|H_z|\leq 0.502645$, corresponding again to an excellent agreement with the specification. The sources of discrepancy are the numerical rounding error and approximations in the averages and differences of the fields at some distance from the metasurface.

\subsection{2D Examples}

The 2D examples to be presented in this section will be compared with COMSOL simulations using a sub-wavelengthly thick metasurface. The synthesis metasurface susceptibility is found by~\eqref{TE}. Since only the $\textrm{TE}_z$ polarization is specified, two of the four susceptibilities can take any values. Therefore, for simplicity but without loss of generalities, we next assume $\chi_{\textrm{me}}^{zy}=\chi_{\textrm{em}}^{yz}=0$. The corresponding metasurface susceptibilities in terms of the specified incident, reflected and transmitted waves are found as
\begin{subequations}
\begin{align}
\chi_{\textrm{mm}}^{zz}=\frac{2}{j\omega\mu_0}\frac{-E_y^{\textrm{tr}}+E_y^{\textrm{ref}}+E_y^{\textrm{inc}}}{H_z^{\textrm{tr}}+H_z^{\textrm{inc}}+H_z^{\textrm{ref}}},\\ \chi_{\textrm{ee}}^{yy}=\frac{2}{j\omega\varepsilon_0}\frac{-H_z^{\textrm{tr}}+H_z^{\textrm{ref}}+H_z^{\textrm{inc}}}{E_y^{\textrm{tr}}+E_y^{\textrm{inc}}+E_y^{\textrm{ref}}}.
\end{align}
\end{subequations}

\begin{figure}[ht]
  \centering
   \subfigure[]{%
   \includegraphics[width=0.5\columnwidth]{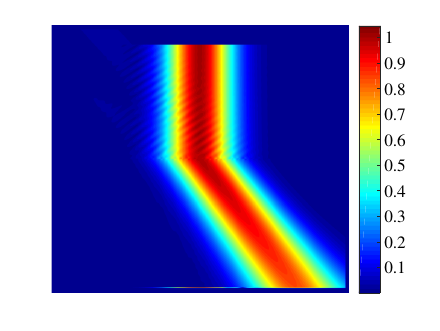}
   \label{FIG:Diff_ABS}}%
   \subfigure[]{%
   \includegraphics[width=0.5\columnwidth]{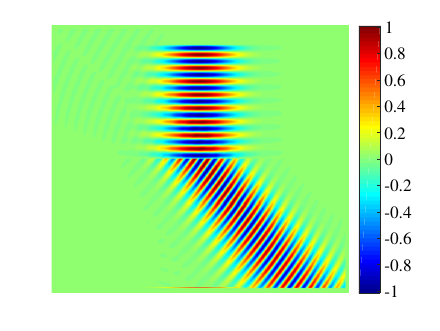}
   \label{FIG:Diff_Im}}%

   \subfigure[]{%
   \includegraphics[width=0.5\columnwidth]{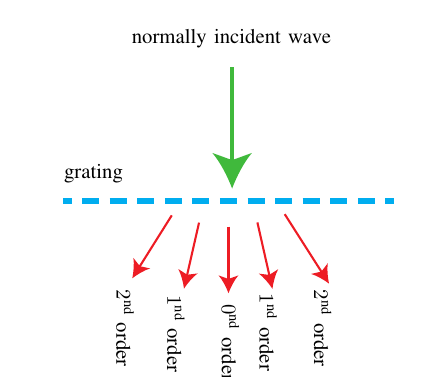}
\label{FIG:diffModes}}%
   \subfigure[]{%
   \includegraphics[width=0.5\columnwidth]{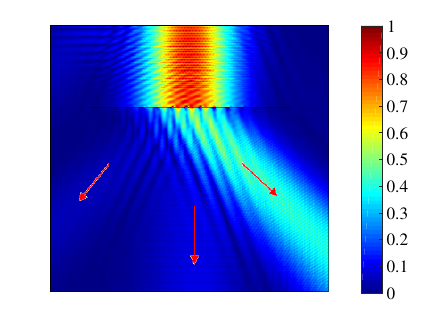}
   \label{FIG:COMSOL}}%
   \caption{Example~3 (2D): FDFD-GSTC results for a refection-less ($R=0$) fully refractive ($T=1$) metasurface deflecting a normally incident ($\theta^\text{inc}=0$) Gaussian beam into an oblique ($\theta^\text{tr}=45^\circ$) Gaussian beam, and comparison with COMSOL. (a)~Absolute value of $H_z$ computed by proposed method. (b)~Imaginary part. (c)~Typical diffraction phenomenon in a periodic structure, such as a grating. (d)~Modeling of the sheet by a thin slab of width $d=\frac{\lambda}{100}$ with 3D susceptibility $\frac{\bar{\bar\chi}}{d}$ by COMSOL.}
   \label{FIG:2DDiff}
\end{figure}

As a first example, consider a reflection-less fully refractive metasurface, whose results are presented in Fig.~\ref{FIG:2DDiff}. The FDFD-GSTC results, plotted in Figs.\ref{FIG:Diff_ABS} and~\ref{FIG:Diff_Im}, perfectly simulate the synthesized metasurface: all the incident wave is transmitted at the specified angle with negligible reflection. The same metasurface, assuming a thickness of $d=\frac{\lambda}{100}$, is simulated using COMSOL, and the result is shown in Fig.~\ref{FIG:COMSOL}. In this case, unspecified refracted beams are clearly visible due to the nonzero thickness of the metasurface. These beams are diffraction orders~\cite{Grating}, represented in Fig.~\ref{FIG:diffModes}, that generally appear in the transmitted region of a grating due to momentum conservation. Such diffraction orders should not appear here because the metasurface was synthesized as a perfect refracting device. While the proposed FDFD-GSTC method satisfactorily simulates this phenomenon, the COMSOL simulation fails to do it because of volume approximation.

\begin{figure}[ht]
  \centering
     \subfigure[]{%
     \includegraphics[width=0.5\columnwidth]{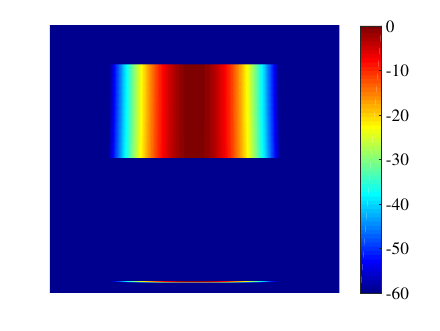}
\label{FIG:ABSabsorber}}%
   \subfigure[]{%
   \includegraphics[width=0.5\columnwidth]{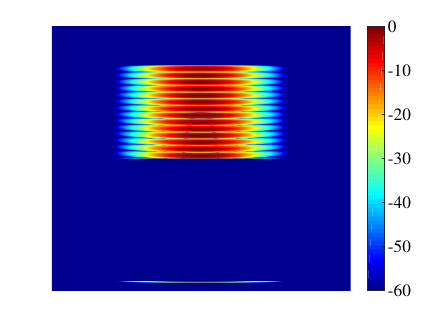}
   \label{FIG:IMabsorber}}%

   \subfigure[]{%
   \includegraphics[width=0.5\columnwidth]{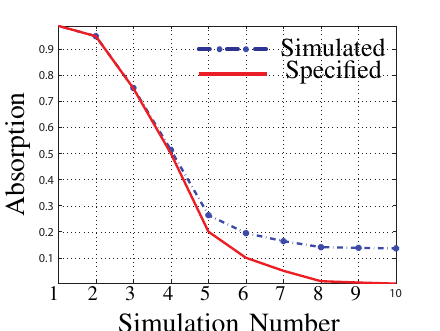}
   \label{FIG:Karim}}%
   \subfigure[]{%
   \includegraphics[width=0.5\columnwidth]{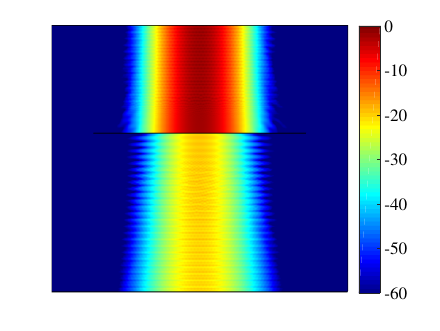}
   \label{FIG:COMSOLAbsorber}}%
   \caption{Example 4 (2D): FDFD-GSTC results for a fully absorptive ($R=T=0$) metasurface, and comparison with COMSOL. (a)~Absolute value of $H_z$ computed by proposed method. (b)~Imaginary part. (c)~Parametric study of COMSOL (same conditions as in Fig.~\ref{FIG:2DDiff}) discrepancy versus absorption level. (d)~COMSOL result for full absorption specification.}\label{Fig:absorber}
\end{figure}

The second 2D example is a fully absorbing metasurface ($R=T=0$) illuminated by a normal gaussian plane wave. The results are shown in Fig.~\ref{FIG:IMabsorber} and~\ref{FIG:ABSabsorber}, in dB for better visualization. Both the reflected and transmitted waves are zero and, therefore all the incident wave has been absorbed by the metasurface in the FDFD-GSTC simulations, shown in Figs.~\ref{FIG:ABSabsorber} and~\ref{FIG:IMabsorber}. The same metasurface is simulated by COMSOL and the result is shown in Fig.~\ref{FIG:COMSOLAbsorber}. It is very well clear that a great amount of the incident wave is parasitically transmitted by the metasurface. In general, as shown in Fig.~\ref{FIG:Karim} and discussed in~\cite{COMSOL}, the COMSOL approach fails to simulate reflection for $T<0.5$. One explanation is the impedance mismatch between the COMSOL's computational slab and the surrounding medium, which results in multiple reflections within the slab and hence transmission of part of the incident wave~\cite{COMSOL}.

\begin{figure}
\centerline{%
 \includegraphics[width=0.5\columnwidth]{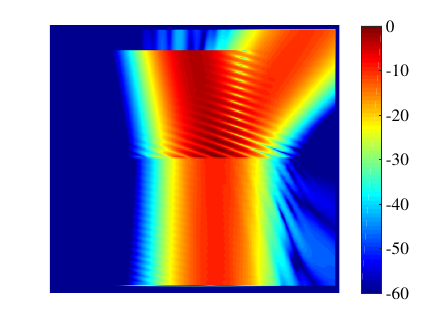}
   %
   \includegraphics[width=0.5\columnwidth]{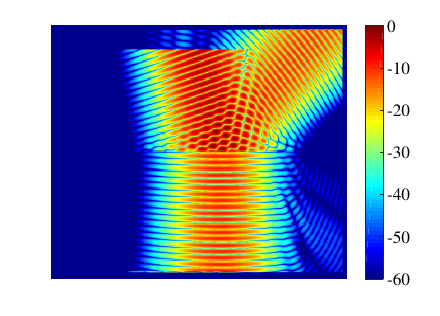}
   %
}%
  \caption{Example 5 (2D): FDFD-GSTC results for a loss-less metasurface equally splitting ($R=T=0.5$) an obliquely incident ($\theta^\text{inc}=15^\circ$) Guassian wave into an obliquely reflected Gaussian wave ($\theta^\text{re}=45^\circ$) and a normally transmitted $\theta^\text{tr}=0$ Gaussian wave. (a)~Absolute value of $H_z$ computed by proposed method. (b)~Imaginary part.}\label{FIG:REFTR}
\end{figure}

The last example is a loss-less partially reflecting and partially transmitting metasurface. The FDFD-GSTC result is shown in Fig.~\ref{FIG:REFTR}, where the reflected and transmitted waves are seen to be in the specified directions. The reflection and transmission coefficients are found, by measurement of maxima, to be $R=0.4964$ and $T=0.497$, respectively. The sources of discrepancy, which is in the order of $-40\textrm{dB}$, is the discretization error and the approximation done in measuring the fields at some distance from the metasurface instead of exactly on the metasurface.

\section{Conclusion}\label{sec:concl}

We have introduced a simple and accurate method for analyzing metasurface discontinuity problems, which represent the most general case of 2D sheet discontinuities. The method has been supported by several illustrative examples and the results  have been advantageously compared with COMSOL approximate solutions. Although the method was presented in the framework of FDFD, it straightforwardly applies to FDTD. This work represents a fundamental contribution to computational electromagnetics since it solves a very canonical and useful problem.

\bibliographystyle{IEEEtran}
\bibliography{TAP_FDFD_SHEET_ALGORITHM_Vahabzadeh}

\appendices
\section{FDFD Implementation}\label{sec:APP1}

For the 1D case, with $n_x$ cells in the $x$ direction, $\ve{D}_\textrm{e}^x$, $\ve{D}_\textrm{h}^x$, $\ve{\varepsilon}_{yy}$ and $\ve{\mu}_{zz}$ are square matrixes of dimension $n_x\times n_x$, while $\ve{E}_y$ and $\ve{H}_z$ are column vectors of $n_x$ rows. By definition, $\ve{D}_\textrm{e}^x\ve{E}_y\equiv\frac{E_y^{i+1}-E_y^{i}}{\Delta{x}}$ or, in matrix form,
\begin{equation}\label{Dez}
\ve{D}_\textrm{e}^x\ve{E}_y=
\frac{1}{\Delta x}
     \begin{tikzpicture}[baseline=(U.center),
                    every left delimiter/.style={xshift=1mm},
                    every right delimiter/.style={xshift=-1mm}]

\matrix [matrix of math nodes,
         inner sep=0pt,
         left delimiter={[},
         right delimiter={]},
         nodes={inner sep=0,minimum size=6mm,anchor=center}
        ] (U) {
-1    & 1     & 0      & 0      & \ldots & 0      \\
0     & -1    & 1      & 0      &\ldots  & 0      \\
0     & 0     & -1     & 1      & \ddots & 0      \\
\vdots&\vdots & \ddots & \ddots & \ddots & \vdots  \\
0     & 0     & \ldots & \ddots & \ddots & 1     \\
0     & 0     & \ldots & 0      & 0      & -1    \\
};
\draw[red] (U-1-1.north) -- (U-1-1.west) -- (U-6-6.south) -- (U-6-6.east) -- cycle;
\draw[red] (U-1-2.north) -- (U-1-2.west) -- (U-5-6.south) -- (U-5-6.east) -- cycle;
\end{tikzpicture}      \begin{tikzpicture}[baseline=(U.center),
                    every left delimiter/.style={xshift=1mm},
                    every right delimiter/.style={xshift=-1mm}]

\matrix [matrix of math nodes,
         inner sep=0pt,
         left delimiter={[},
         right delimiter={]},
         nodes={inner sep=0,minimum size=6mm,anchor=center}
        ] (U) {
E_y^1\\
E_y^2\\
E_y^3\\
\vdots\\
E_y^{n_z-1}\\
E_y^{n_z}\\
};
\end{tikzpicture}
\end{equation}
and
\begin{equation}\label{muhz}
    \ve{\mu}_{zz}\ve{H}_z=
    \begin{tikzpicture}[baseline=(U.center),
                    every left delimiter/.style={xshift=1mm},
                    every right delimiter/.style={xshift=-1mm}]

\matrix [matrix of math nodes,
         inner sep=0pt,
         left delimiter={[},
         right delimiter={]},
         nodes={inner sep=0,minimum size=6mm,anchor=center}
        ] (U) {
\alpha    & 0    & 0      & 0      & \ldots & 0      \\
0     & \alpha    & 0      & 0      &\ldots  & 0      \\
0     & 0     & \alpha     & 0      & \ddots & 0      \\
\vdots&\vdots & \ddots & \ddots & \ddots & \vdots  \\
0     & 0     & \ldots & \ddots & \ddots & 0     \\
0     & 0     & \ldots & 0      & 0      & \alpha\\
};
\draw[red] (U-1-1.north) -- (U-1-1.west) -- (U-6-6.south) -- (U-6-6.east) -- cycle;
\end{tikzpicture}     \begin{tikzpicture}[baseline=(U.center),
                    every left delimiter/.style={xshift=1mm},
                    every right delimiter/.style={xshift=-1mm}]

\matrix [matrix of math nodes,
         inner sep=0pt,
         left delimiter={[},
         right delimiter={]},
         nodes={inner sep=0,minimum size=6mm,anchor=center}
        ] (U) {
H_z^1\\
H_z^2\\
H_z^3\\
\vdots\\
H_z^{n_z-1}\\
H_z^{n_z}\\
};
\end{tikzpicture}
\end{equation}
where $\alpha=-j\omega\mu_0\mu_{zz}$. Similarily, $\ve{D}_{\textrm{h}}^x\ve{H}_z\equiv\frac{H_z^{i+\frac{1}{2}}-H_z^{i-\frac{1}{2}}}{\Delta{x}}$ can be cast as
\begin{equation}\label{Dhz}
\ve{D}_{\textrm{h}}^x\ve{H}_z=\frac{1}{\Delta x}
  \begin{tikzpicture}[baseline=(U.center),
                    every left delimiter/.style={xshift=1mm},
                    every right delimiter/.style={xshift=-1mm}]

\matrix [matrix of math nodes,
         inner sep=0pt,
         left delimiter={[},
         right delimiter={]},
         nodes={inner sep=0,minimum size=6mm,anchor=center}
        ] (U) {
1    & 0     & 0      & \ldots      & 0 & 0      \\
-1     & 1    & 0      & \ldots      & 0 & 0      \\
0     & -1     & 1     & \ldots      & 0 & 0      \\
\vdots&\vdots & \ddots & \ddots & \ddots & \vdots  \\
0     & 0     & \ldots & -1 & 1 & 0     \\
0     & 0     & \ldots & 0      & -1      & 1\\
};
\draw[red] (U-1-1.north) -- (U-1-1.west) -- (U-6-6.south) -- (U-6-6.east) -- cycle;
\draw[red] (U-2-1.north) -- (U-2-1.west) -- (U-6-5.south) -- (U-6-5.east) -- cycle;
\end{tikzpicture}   \begin{tikzpicture}[baseline=(U.center),
                    every left delimiter/.style={xshift=1mm},
                    every right delimiter/.style={xshift=-1mm}]

\matrix [matrix of math nodes,
         inner sep=0pt,
         left delimiter={[},
         right delimiter={]},
         nodes={inner sep=0,minimum size=6mm,anchor=center}
        ] (U) {
H_z^1\\
H_z^2\\
H_z^3 \\
\vdots\\
H_z^{n_z-1}\\
H_z^{n_z}\\
};
\end{tikzpicture}
\end{equation}
and the matrix form $\ve{\varepsilon}_{yy}\ve{E}_y$ is the same as $\ve{\mu}_{zz}\ve{H}_{z}$ in~\eqref{muhz} but with $\alpha=j\omega\varepsilon_0\varepsilon_{yy}$ and $H_z$ replaced by $E_y$.

For the 2D problem, the operators $\ve{D}_\textrm{e}^x$ and $\ve{D}_{\textrm{h}}^z$, the matrices $\ve{\varepsilon}_{yy}$ and $\ve{\mu}_{zz}$, and the vectors $\ve{H}_z$ and $\ve{E}_y$ only slightly differ from their 1D counterpart. Assuming $n_x$ and $n_y$ cells in the $x$ and $y$ directions, respectively, all dimensions $n_x$ will increase to $N=n_xn_y$. We will have for instance $\ve{D}_{\textrm{h}}^y\ve{H}_z\equiv \frac{H_z^{i+\frac{1}{2},j+\frac{1}{2}}-H_z^{i+\frac{1}{2},j-\frac{1}{2}}}{\Delta y}$ reading
\begin{equation}\label{DeyDhy}
\ve{D}_{\textrm{h}}^y\ve{H}_z=
\frac{1}{\Delta y}
  \begin{tikzpicture}[baseline=(U.center),
                    every left delimiter/.style={xshift=1mm},
                    every right delimiter/.style={xshift=-1mm}]

\matrix [matrix of math nodes,
         inner sep=0pt,
         left delimiter={[},
         right delimiter={]},
         nodes={inner sep=0,minimum size=6mm,anchor=center}]
         (U) {
 -1    & 0     &\ldots& 1      & 0    &\ldots &0 \\
 0     &-1     &    0 &\ddots  &   1  &\ddots &0  \\
 0     & 0     &\ddots& 0      &\ddots&\ddots &0 \\
 \vdots& \ddots& \ddots&-1     &\ddots &   0   &1 \\
 0     & \ddots& \ddots&\ddots & \ddots&\vdots &\vdots\\
 0     &    0  & \ldots&\ldots &0   &  -1   & 0\\
 0     & 0     & 0     &\ldots& \ldots&0      & -1\\
};
\draw[red] (U-1-1.north) -- (U-1-1.west) -- (U-7-7.south) -- (U-7-7.east) -- cycle;
\draw[red] (U-1-4.north) -- (U-1-4.west) -- (U-4-7.south) -- (U-4-7.east) -- cycle;
\end{tikzpicture}   \begin{tikzpicture}[baseline=(U.center),
                    every left delimiter/.style={xshift=1mm},
                    every right delimiter/.style={xshift=-1mm}]

\matrix [matrix of math nodes,
         inner sep=0pt,
         left delimiter={[},
         right delimiter={]},
         nodes={inner sep=0,minimum size=6mm,anchor=center}]
         (U) {
 H_z^1   \\
 H_z^2     \\
 H_z^3      \\
 \vdots \\
 H_z^{N-2}     \\
 H_z^{N-1}     \\
 H_z^{N}     \\
};
\end{tikzpicture}
\end{equation}
where the $-1$ and $1$ entries in row $i$ are placed in the columns $i$ and $i+n_x$, respectively. The matrices and vectors are easily deduced from this.
\end{document}